\documentclass[twocolumn,aps,amsmath,amssymb]{revtex4-1}
\usepackage{amssymb}
\usepackage{amsmath}
\usepackage{graphicx}
\usepackage{epstopdf}
\usepackage{bm}%
\usepackage{gensymb}
\usepackage{soul}
\usepackage{sidecap}
\usepackage[normalem]{ulem}
\usepackage{epsfig}
\usepackage{graphicx}
\usepackage{amsfonts,amssymb}
\usepackage{amsmath}
\usepackage{color}
\usepackage {times}
\usepackage{epstopdf}

\newcommand{\bl}{\begin{aligned}}
\newcommand{\el}{\end{aligned}}

\newcommand{\be}{\begin{equation}}
\newcommand{\ee}{\end{equation}}   
\newcommand{\bea}{\begin{eqnarray}}
\newcommand{\eea}{\end{eqnarray}}
\newcommand{\ba}{\begin{array}}
\newcommand{\ea}{\end{array}}

\newcommand{\q}{{\bf q}}
\renewcommand{\k}{{\bf k}}
\newcommand{\Q}{{\bf Q}}

\graphicspath{{Figs/}}

\begin{document}
\title{Quasi-one Dimensional Nanostructures as Sign of Nematicity in Iron Pnictides and Chalcogenides}
\date{\today}
\author{Dheeraj Kumar Singh$^{1,2,3}$}\email{dheeraj@postech.ac.kr }
\author{Alireza Akbari$^{2,3,4}$}\email{alireza@apctp.org}
\author{Pinaki Majumdar$^1$}
\affiliation{$^1$Harish-Chandra Research Institute, HBNI,  Chhatnag Road, Jhunsi, Allahabad 211019, India}
\affiliation{$^2$Asia Pacific Center for Theoretical Physics, Pohang, Gyeongbuk 790-784, Korea}
\affiliation{$^3$Department of Physics, POSTECH, Pohang, Gyeongbuk 790-784, Korea}
\affiliation{$^4$Max Planck POSTECH/Korea Research Initiative (MPK), Gyeongbuk 376-73, Korea }
%
\begin{abstract}
Impurity scattering is found to lead to quasi-one dimensional nanoscale 
modulation of the local density of states in the iron pnictides and 
chalcogenides. This `quasiparticle interference' feature  is remarkably 
similar across a wide variety of pnictide and chalcogenide phases, suggesting 
a common origin. 
 We show that a unified understanding of the experiments can 
be obtained by simply invoking a four-fold symmetry breaking $d_{xz}-d_{yz}$ 
orbital splitting, of a magnitude already suggested by the experiments. 
This can explain the one-dimensional characteristics in the local density 
of states observed in the orthorhombic nematic, tetragonal paramagnetic, as well as the 
spin-density wave and superconducting states in these materials.
\end{abstract}
\maketitle

The intriguing anisotropic electronic properties of iron pnictides~\cite{stewart} are reflected in transport measurements~\cite{chu, tanatar, blomberg}, 
optical conductivity~\cite{nakajima}, angle-resolved photoemission 
spectroscopy (ARPES)~\cite{yi}, and scanning tunneling microscopy (STM)
\cite{chuang}. 
It is not unexpected in 
a state having a broken four-fold rotational symmetry such as the 
spin-density wave (SDW) state or the orthorhombic `spin nematic' state,
 but the lattice anisotropy does not explain the 
splitting of $\approx$ 60meV between the $d_{xz}$ and
$d_{yz}$ orbitals~\cite{kontani,kasahara}. The orbital splitting (OS) actually 
persists into the high temperature tetragonal phase~\cite{kasahara}. 
This suggests that the OS, rather than the orthorhombic
symmetry or magnetic order, could be 
the key player in electronic anisotropy.
A similar OS exists in various phases 
\cite{Song:2011aa, song,shimojima,nakayama,watson, baek} 
of the chalcogenide including the
superconducting state. The energy scale of
FeSe splitting, and its orbital character, 
has been contrasted  with those of the 
pnictides, with some suggestions of 
a momentum dependent, {\it i.e}, non-uniform splitting. Unlike the pnictides 
where the degeneracy of bands is dominated mainly by $d_{xz}$ and $d_{yz}$ orbitals 
at X or Y points 
is lifted at low temperature, the OS for chalcogenides may also
exhibit sign reversal. Some have reported it to be of entirely 
different nature, OS 
between $d_{{xz}/{yz}}$ and $d_{xy}$~\cite{zhang2,suzuki}.

Valuable insight into electronic 
anisotropy can be obtained through the 
`quasiparticle interference' (QPI) phenomena which
basically probes the spatial variation of the local
density of states (LDOS), due to impurities in the medium,
using the spectroscopic imaging 
STM~\cite{hoffman,*huang,*Hanaguri:2010aa, *Hirschfeld:2015aa,*Sprau:2017aa,*Boker:2017aa,*Singh:2017aa,*Martiny:2017aa,*Altenfeld:2018aa,*Kamble:2016aa,*Du:2017aa,*Choi:2017aa,*Sykora:2011aa,*Singh:2017ab,*Kostin:2018aa}. 
A remarkable characteristic of the QPI common to the SDW state, the orthorhombic nematic phase, and the 
tetragonal paramagnetic phase, in some of the pnictides is the occurrence of quasi-one dimensional real-space LDOS modulation with material dependent
  lengthscale~\cite{ allan,chuang,rosenthal}. 
  Corresponding momentum-space structure in the form of almost parallel ridges are aligned along a direction 
  reciprocal to the ferromagnetic direction in the SDW state, or $b$-axis in the orthorhombic phase for pnictides. Similar momentum- and real-space
   structures have been reported in superconducting phase of  chalcogenides~\cite{song}. This suggests a common origin of the anisotropy in the electronic structure, rather than in  specific ordering tendencies.

 In the SDW state, 
the orbital occupancy difference that can 
result from the electronic reconstruction 
is $n_{xz} - n_{yz} \sim 0.1$~\cite{bascone}, 
which corresponds roughly to an energy 
splitting of $50$meV. 
According to the experiments, the OS observed above Neel temperature 
$T_N$ can be as large as $\sim  60$meV~\cite{yi}, 
therefore it is natural to explore the consequences of
this `orbital bias' 
in studying the SDW state as well,   
ignored in earlier work which may have led to their failure in reproducing the one dimensional (1d) characteristics with correct orientation and lengthscale~\cite{knolle,*akbari,mazin,plonka,zhang}. 
Such a term should assume further importance, 
beyond magnetic anisotropy,
in the electron-doped region of SDW state 
where the magnetic moments are small, and 
magnetic order induced band reconstruction
is less pronounced.

Above $T_N$, a non-zero OS has been attributed to the spin-driven nematic order with $\langle {\bf S}_{\bf i}\cdot {\bf S}_{\bf i + x}  -{\bf S}_{\bf i}\cdot {\bf S}_{\bf i + y} \rangle \ne 0$, where average magnetic moment $\langle {\bf S}_{\bf i}\rangle = 0$, because of the frustration caused by the presence of second nearest-neighbor exchange coupling~\cite{fang,Fernandes:2012aa,Fernandes:2014aa,*fernandes}. It is not clear enough how this mechanism will support OS term of similar strength below $T_N$ in the SDW state, which we find necessary to explain the 1d QPI characteristics. 
In another scenario, OS may also originate from the ferro-orbital order~\cite{kruger,lv, *Lv:2010aa,lee,kontani1,*Kontani:2014aa,*Yamakawa:2016aa} caused by the spin-orbital mode coupling, which can be responsible for an OS larger than what is expected merely induced by the SDW state. Behavior of orbital order appears to have a remarkable similarity to some of the manganites where the orbital order precedes the magnetic order as temperature is lowered~\cite{dhesi} except that the lattice distortion is small enough in iron-based superconductors to account for such a large OS.

In this letter, we suggest a unified explanation for the  
common QPI characteristics of different phases of iron-based 
superconducting systems. Our proposal is that an explicit OS term 
in the Hamiltonian is crucial irrespective of phases. Thus, our point of departure in the standard five-orbital Hamiltonian is the 
OS term:
\be
{\cal H}_{orb}\!\!=\!-\frac{\delta}{2} \sum_{i \sigma}(d^{\dagger}
_{i  xz \sigma}d^{}_{i  xz \sigma}
\!-\! d^{\dagger}_{i  yz \sigma}d^{}_{i  yz \sigma})
.
\ee
Here, 
$d^{\dagger}_{i  \gamma \sigma} (d^{}_{i  \gamma \sigma})$
 is 
the creation (annihilation) operator 
for an electron in the $d_{\gamma}$-orbital  with spin $\sigma$ at site $i$. 
The impurity scattering effects that generate the spatial 
LDOS modulations,
{\it i.e}, QPI patterns, are handled via a $t$-matrix approach on 
the mean field states of this theory.
\\

\begin{figure}[t]
\begin{center}
\psfig{figure=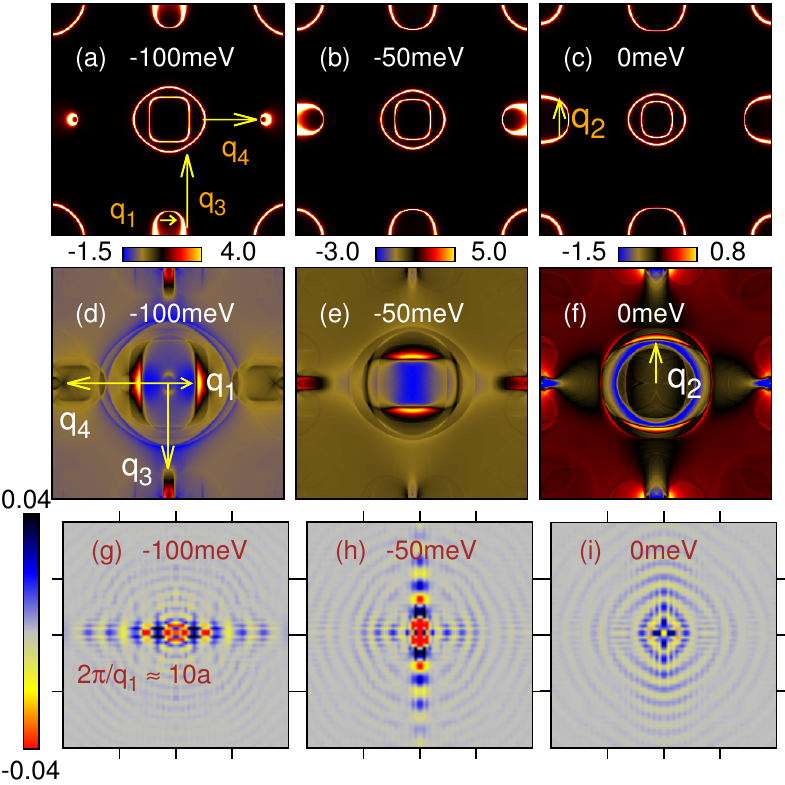,width=1 \linewidth}
\end{center}
\vspace{-3mm}
\caption{
Results in the nematic state for 
orbital splitting $\delta = 60$ meV:
(a-c) show the behavior of contours of constant energies (CCEs)
for the quasiparticle energy $\omega=-100,-50,0$meV
in the ($k_x$, $k_y$)-plane. $\q_1$ and $\q_2$ are 
intrapocket scattering vectors associated with the electron 
pockets around (0, $\pm \pi$) and ($\pm \pi$, 0), respectively. 
Intrapocket scattering vectors for the hole pockets around (0,0) are not shown. $\q_3$ and $\q_4$ are the interpocket scattering vectors.
(d-f) For most $\omega$ three parallel rod-like
structures exist in the momentum space QPI, 
the outer peaks are positive the inner peak is negative.
Since the orientation of these rod-like
structures also changes near
$\omega$ $\sim$ -60meV, orientation of 1d LDOS modulation (g-i) also changes from $x$ to $y$.
Note: here and hereafter the momentum   space plots  are in the units of $\pi/a$ with   
range [-1, 1]; 
and the real space  ($xy$-plane) plots are in the units of $a$ with range [-40, 40].
LDOS modulation shown for 80$\times$80 size with
the impurity atom located at the center, calculation  done for 
300$\times$300 lattice size.
}
\label{qpi1}
\end{figure}

\begin{figure}[t]
\begin{center}
\psfig{figure=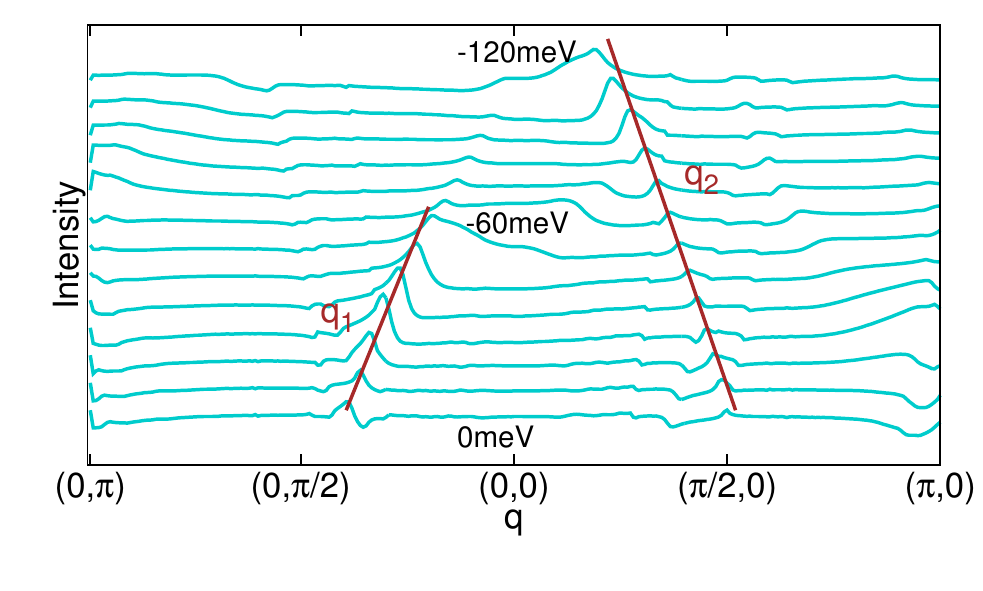,width=1 \linewidth}
\end{center}
\vspace{-12mm}
\caption{ 
QPI along the high-symmetry directions in the nematic state, for different energies, from $\omega = 0.0$meV  (bottom curve) to  
$\omega = -120$meV (top curve) with energy step of $10$meV. 
The brown and red curves are guide to the eye for scattering vectors $\q_1$ and $\q_2$.
}
\label{disp1}
\end{figure}

Our key results are listed as the followings:
(i)~We obtain nearly 
1d LDOS modulations,
{\it i.e}, real-space QPI patterns, 
a feature observed universally across various phases. 
(ii)~For the five-orbital model used in this work, the 
wavelength of  the 
 modulations is 
$~\sim 8 a_{\rm Fe-Fe}$ in excellent
agreement with STM measurements for the SDW state of 
Ca(Fe$_{1-x}$Co$_x$)$_2$As$_2$~\cite{chuang}.  
(iii)~We identify two large energy windows of size $\sim 60$meV where the LDOS modulation is one dimensional. 
For the SDW state it is oriented along the antiferromagnetic   direction as observed in the experiment.
This happens when the energy of the $d_{xz}$ orbital is lower than that of $d_{yz}$.
(iv)~The key factor responsible for all the findings above is the OS term which
leads to the upward or downward shift of
either set of electron pockets located around 
($0, \pm \pi$) or ($\pm \pi, 0$). Combined with a large
spectral density due to nearby band extrema, it
results in a strongly momentum dependent 
spectral density along 
the constant energy contours, yielding 
the anisotropic patterns.
\\

We start to analyze  the  QPI in the
superconducting (SC) phase.  The mean 
field (MF) Hamiltonian written in the Nambu formalism is
%
\bea
{\cal H}_{sc} = \sum_{\k} 
\Psi^{\dagger}_{\k}
\begin{pmatrix}
\hat{\varepsilon}_{\k} & \hat{\Delta}_{\k} \\
\hat{\Delta}^{\dagger}_{\k} & -\hat{\varepsilon}_{\k}
\end{pmatrix}
 \Psi_{\k} ,
\eea
%
where the electron field operator is defined as 
$\Psi^{\dagger}_{\k } = (\phi_{{\bf k}\uparrow}^{\dagger},\phi_{-{\bf k}\downarrow}^{})$
with 
$\phi_{{\bf k}\uparrow}^{\dagger}= (d^{\dagger}_{{\bf k}1\uparrow}, \cdots,d^{\dagger}_{{\bf k}5\uparrow})$
where 
subscript 1 to 5 denoting the five $d$ orbitals 
$d_{3z^2-r^2}$, $d_{xz}$, $d_{yz}$, $d_{x^2-y^2}$, and $d_{xy}$ 
in the same order. 
Here, $\hat{\varepsilon}_{\k}$ is a 5$\times$5 hopping matrix~\cite{ikeda}, and
 $\hat{\Delta}_{\k}$ is a 5$\times$5 pairing matrix.
Effective $s^{+-}$~pairing 
state is mediated by the antiferromagnetic fluctuations generated by 
the interplay of Fermi surface nesting and on site Coulomb interaction, and the interaction  part of the Hamiltonian  is given~by
\begin{eqnarray}
\hspace{-0.5cm}
\bl
{\mathcal H}_{int} 
= \;
& U \sum_{{\bf i},\mu} n_{{\bf i}\mu 
\uparrow} n_{{\bf i}\mu \downarrow} + (U' -
\frac{J}{2}) \sum_{{\bf i}, \mu<\nu} n_{{\bf i} \mu} 
n_{{\bf i} \nu} 
 \\ 
&- 
2 J 
\!\!
\sum_{{\bf i}, \mu<\nu} {\bf{S_{{\bf i} \mu}}} 
\cdot {\bf{S_{{\bf i} \nu}}} 
+ J' 
\!\!\!\!
\sum_{{\bf i}, \mu<\nu, \sigma} 
\!\!\!\!
d_{{\bf i} \mu \sigma}^{\dagger}d_{{\bf i} 
\mu \bar{\sigma}}^{\dagger}d_{{\bf i} \nu \bar{\sigma}}
d_{{\bf i} \nu \sigma}. 
\label{int}
\el
\end{eqnarray}
Here, the respective terms represent intraorbital, 
interorbital density-density, Hund's coupling and pair-hopping 
energy ($J'=J$)  in the given order.
For simplicity,   
we consider only intra-orbital pairing with 
the  same and isotropic  gap elements, ${\Delta}_0 \cos k_x
\cos k_y$, (in general, the SC gap is also expected to be 
anisotropic~\cite{liu}). We set ${\Delta}_0=20$meV and  the bandfilling is fixed at $n= 6.1$.
The complete Hamiltonian is given by ${\cal H}_{orb} + {\cal H}_{sc} + {\cal H}_{\rm imp} $, where
${\cal H}_{\rm imp}$ = 
$\sum_{\mu \sigma} V_{\rm imp} d_{j \mu \sigma}^{\dagger} 
d_{j \mu \sigma}^{}$
accounts for a non-magnetic delta like impurity
scatterer present at site $j$. 
The 
modulation caused in the LDOS by the impurity term is 
calculated within the $t$-matrix approximation
and only orbitally diagonal scattering 
is retained~\cite{zhang1}. 
\\%

The MF Hamiltonian in the  SDW state is obtained after standard 
decoupling of the on-site terms in Eq.~(\ref{int}) as 
\be
  {\mathcal{H}}_{SDW}  = 
\sum_{{\bf k }\sigma}\Psi'^{\dagger}_{{\bf k} \sigma}
\begin{pmatrix}
{\hat{\varepsilon}_{\k}+\hat{N}}  & {{\rm sgn}\bar{\sigma}\hat{W}} \\
{{\rm sgn}\bar{\sigma}\hat{W}} & {\hat{\varepsilon}_{\bf {k+Q}}+\hat{N}}
\end{pmatrix}
\Psi'_{{\bf k} \sigma}.
\ee
%
Here the new electron field operator is defined as $\Psi'^{\dagger}_{{\bf k} \sigma} = (\phi_{{\bf k}\sigma}^{\dagger},\phi_{{\bf k}+{\bf Q}\sigma}^{\dagger})$  
 with the ordering wavevector $\Q = (\pi,0)$.
Matrices $\hat{N}$ and $\hat{W}$ are obtained 
in a self-consistent manner. The bandfilling in this case is $n = 6.0$.
We chose intraorbital Coulomb interaction 
$U = 0.96$eV and Hund's coupling $J = 0.25U$ while 
pair-hopping interaction $J^{\prime} = J$ and 
interorbital density-density interaction parameter 
$U^{\prime} = U-2J$ are determined by the standard relations. 
Our choice of $U$ yields a net magnetization $m = 0.3$ consistent 
with the experiments~\cite{cruz}.
\\

\begin{figure}[t]
\begin{center}
\hspace{-4.mm}
\psfig{figure=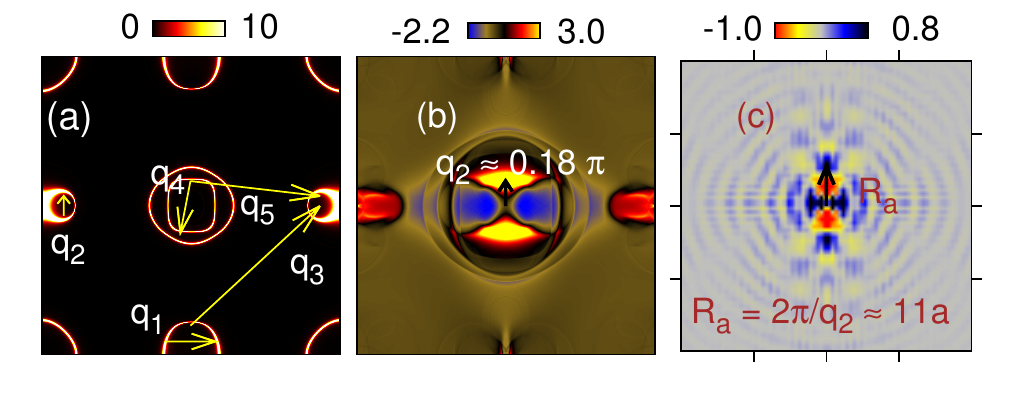,width=1.045 \linewidth}
\end{center}
\vspace{-6mm}
\caption{
Results in the  $s^{+-}$ superconducting state for $\Delta_0 = 20$meV, and  quasiparticle energy $\omega = -88$meV  in the presence of orbital splitting $\delta = 60$meV:
(a) Quasiparticle spectral function
in ($k_x$, $k_y$)-plane,
(b) Momentum space QPI 
in ($q_x$, $q_y$)-plane, and 
(c) Real-space QPI. 
The  pockets in (a) at ($\pm\pi, 0$)
are very small with a highly anisotropic
spectral density distribution along them.
The intrapocket scattering vector
$\q_2$ is mainly responsible for
the features observed near (0, 0) in
the momentum-space QPI pattern. These 
consist of three parallel rod-like 
structures, the outer ones with positive peak
and inner one with negative peak. 
Real-space
QPI consists of two bright spots separated by a distance of
$\sim 11a_{{\rm Fe}-{\rm Fe}}$ as observed in
 the experiments~\cite{uday}.
 Range for all the quantities are
as in Fig.~\ref{qpi1}.
 }
\label{qpi2}
\end{figure}

Now we discuss the QPI results. Throughout, the impurity potential strength   is set to be  $V_{\rm imp}=200$meV, and the mesh size of 300$\times$300 in the
momentum space is used. Real-space QPI or LDOS modulation is 
obtained using the property of Fourier transform. We set the OS  to be
$\delta = 60$meV unless stated otherwise. 
QPI in the nematic phase
is calculated by setting 
the SC order parameters to zero  with bandfilling $n$ = 6.0 (for different OS values see supplementary).

To understand
QPI patterns in the orthorhombic nematic or
tetragonal paramagnetic phase as shown in
Fig.~\ref{qpi1}(d-i), we first examine the
quasiparticle spectral functions 
[Fig.~\ref{qpi1}(a-c)].  An important consequence
of a non zero $\delta$ is the difference in size
of the two sets of pockets around ($\pm \pi, 0$) and ($0, \pm \pi$) 
with large but non-uniform spectral density 
along both of them (see Fig.~\ref{qpi1}(a)). Note that 
the pockets are on the verge of disappearance near  $\omega \sim -100$meV. 
The spectral density is larger along these pockets because of the nearby 
extrema. As a result, $\q_1$ and $\q_2$ are the
important scattering vectors, and among them 
 those 
aligned parallel to either of $x$- or $y$- directions 
are the most prominent ones, as they connect the regions 
dominated by the same orbital. This follows
straight from the fact that only 
intraorbital scattering is allowed. The main
consequence to be described below is the orientation of LDOS modulation 
along either $x$ or $y$. 

For $\omega \sim -100$meV, $\q_1$ associated with the 
electron pockets around ($0, \pm \pi$) 
should be the dominant  scattering vector
despite the fact that $\q_2$ does also connect the 
pockets having larger spectral density. That is because of
the availability of a larger phase space as the electron pockets are bigger in 
contrast with those around ($ \pm \pi$, 0). In particular, $\q_1$s which are
parallel to $x$-direction  should dominate the QPI patterns. 

\begin{figure}[b]
\begin{center}
\psfig{figure=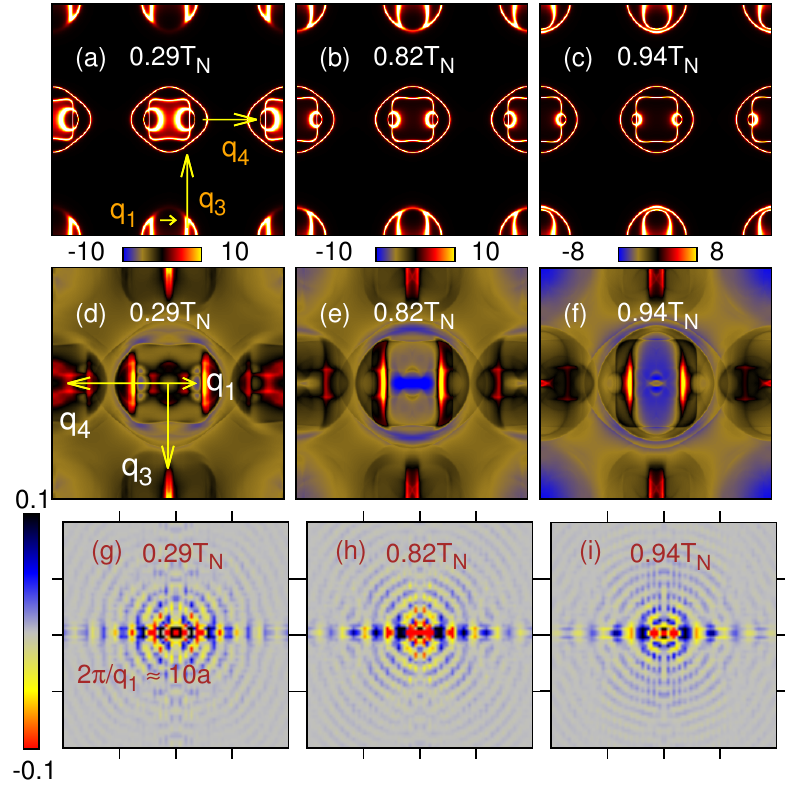,width=1 \linewidth}
\end{center}
\vspace{-4mm}
\caption{Panels (a-c) show the quasiparticle
spectral function for $-100$meV in the ($\pi,0$) SDW
state for various temperatures. Total magnetization is
$m_{\rm tot} = 0.3$. The resulting momentum-space pattern with
three parallel rod-like structures is along a direction reciprocal
to ferromagnetic chain and
LDOS modulation with wavelength $\sim 8-10a$ 
 with   the  small change in temperature.
Range for all the quantities are
as in Fig.~\ref{qpi1}.}
\label{qpi3}
\end{figure}
\begin{figure}[t]
\begin{center}
\psfig{figure=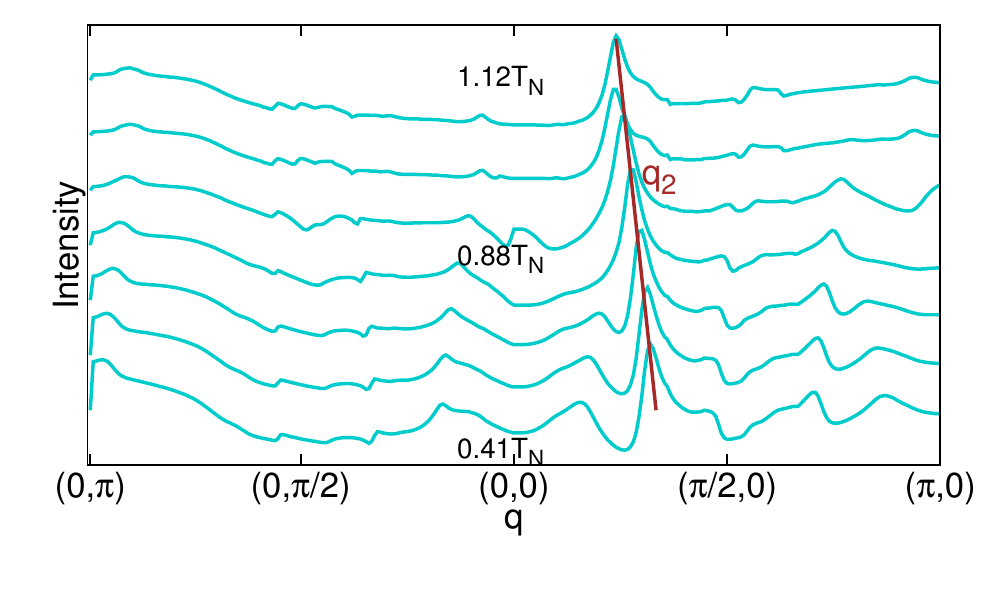,width=1 \linewidth}
\end{center}
\vspace{-12mm}
\caption{QPI along the high-symmetry directions
in the SDW state for different temperatures, starting from $T = 0.014t$  (bottom curve) to $T = 0.036t$ (top curve) with step of $0.04t$
(T$_N$ $\approx$ 0.034$t$).  
 The brown curve is
a guide to the eye for scattering vector $\q_2$.}
\label{disp2}
\end{figure}
When energy increases through $\omega \sim -60$meV, 
contours of constant energies (CCEs) move away from the band extrema, and the smaller pocket 
around ($\pm \pi$, 0) grows
while the bigger ones around ($\pm \pi$, 0) do not show 
much change. However, the spectral density 
along the pockets around ($\pm \pi$, 0) becomes  larger 
in comparison to that along the pocket
around ($0, \pm \pi$). Thus, $\q_2$ instead of $\q_1$ 
is now  the dominant scattering vector.
CCEs move further away from the band extrema, when 
$\omega$ increases  and crosses
$\sim 0$meV. Then, the 
QPI patterns are expected to 
become nearly isotropic and featureless. 

As anticipated, a larger spectral density 
along the sides parallel to the major axis of
elliptical CCEs around ($\pm \pi$, 0) and 
($0, \pm \pi$) results in the dominance of $\q_1$ or
$\q_2$ in the momentum-space QPI patterns, 
which is shown in Fig.~\ref{qpi1}(d-f). When $\omega
< -60$meV, $\q_1$ leads to a nearly parallel rod-like positive peak 
structures at $\sim (\pm \pi/5, 0)$ running
parallel to $q_x \sim \pm \pi/5$. A negative peak 
structure along $q_x \sim 0$ is also seen.
When $\omega$ decreases and crosses $-60$meV, $\q_2$ 
instead of $\q_1$ becomes relevant and the 
patterns are rotated by 90$^\circ$. Near $\omega = 0$, 
QPI is featureless. A recent SI-STM on FeSe$_{1-x}$S$_x$ does also 
report an isotropic QPI patterns for positive $\omega$~\cite{hanaguri}. 

Figs.~\ref{qpi1}(g-i) show the real-space
QPI in the immediate vicinity of 
the impurity atom on a 80$\times$80 lattice 
size for better visibility though the calculation was 
done for 300$\times$300 lattice size. Nearly 1d LDOS modulation is 
obtained over a wide energy window of $\sim 100$meV 
centered around $\omega \approx -60$ meV. As expected,
modulating directions are orthogonal 
to each other $i.e.$ along $x$ and $y$ for $\omega\lesssim 
-60$meV and $\omega \gtrsim -60$meV, respectively. The wavelength 
of modulation for $\omega = -100$meV is
$\lambda_n \sim 10a_{{\rm Fe}-{\rm Fe}}$, which 
is close to $\sim$ 13$a_{{\rm Fe}-{\rm Fe}}$ 
observed in the nematic state of NaFeAs 
\cite{rosenthal}. Note that QPI dispersion 
shows an almost linear dependence for 
$\q_1$s and $\q_2$s, which are centered 
around (0, $ \pi/4$) and ($\pi/4$, 0), respectively [Fig.~\ref{disp1}].

Figs.~\ref{qpi2}(a,b) show calculated
quasiparticle spectral function
and QPI in the
momentum space  in the SC state for energy
$\omega = -88$meV, respectively. The LDOS modulation obtained
by using the Fourier transform is shown in Fig.\ref{qpi2}(c).
As can be seen, a nanostructure in
the vicinity of impurity atom centered around
(0, 0) exists with orientation along $y$, which can change with
energy to $x$. 
The distance between two consecutive 
bright spots is $\sim11 a_{{\rm Fe}-{\rm Fe}}$. STM measurement in the SC state of FeSe$_{1-x}$S$_x$ with broken four-fold rotation symmetry reports scattering vector $q_x \sim \pi/8$~\cite{hanaguri}. Similar scattering vectors have been reported earlier in FeSe as well FeSe$_{0.4}$Te$_{0.6}$~\cite{uday}.
 Thus, our results show good agreement with the experiments. Dependence of QPI pattern on the quasiparticle energy is similar in various aspects to that in the nematic state.

QPI patterns obtained for  energy
$\omega=-100$meV in the SDW state as shown in
Fig.~\ref{qpi3}(d-i) 
are the central results of this work.
The dominant effect of the OS term 
can  be  easily seen even-though there is significant reconstruction of the  band-structure. 
The 
band retains the salient features of nematic state and show only 
a little change with temperature.
Consequently, the LDOS modulation is nearly 1d with 
orientation along the antiferromagnetic direction Figs.~\ref{qpi3}(g-i). 
Similarly, the momentum-space QPI patterns consist of 
parallel running peak structures in a direction reciprocal to 
the ferromagnetic direction. An additional negative peak 
structure is present in between the two. All of 
these characteristics are in excellent agreement the STM results including a small change  with the  temperature.
The later can be seen from the fact that there is only a little change in the scattering vector magnitude
as a function of quasiparticle energy [see Fig.~\ref{disp2}].

Similar QPI patterns have been observed in the 
SDW state of Ca(Fe$_{1-x}$Co$_x$)$_2$As$_2$
\cite{chuang,allan} and NaFeAs~\cite{rosenthal}. The wavelengths for 1d LDOS modulation in the two pnictides are $\approx$ 8$a_{{\rm Fe}-{\rm Fe}}$ and $\approx$ 13$a_{{\rm Fe}-{\rm Fe}}$, respectively, which compares well with $\sim 8a_{{\rm Fe}-{\rm Fe}}$ obtained within the five-orbital of Ref.~\cite{ikeda} considered in this work (see the TABLE~I).
%
\begin{table}[h]
\caption{Size of one-dimensional nanostructures in the LDOS modulation of various iron-based superconductors. We obtain the length scale of the LDOS modulation  $\sim 6a-11a$.} 
\centering 
\setlength{\tabcolsep}{3pt} 
    \renewcommand{\arraystretch}{1} 
\begin{tabular}{c|c c c c c c c c c c c c c c c c c c c c c c} 
\hline\hline 
 Phase  & Nematic & SDW & SC \\ [0.5ex] 
 \hline
Ca(Fe$_{1-x}$Co$_x$)$_2$As$_2$~\cite{chuang,allan} &$ - $ & 8$a$ &  $-$ & \\
\hline
NaFeAs~\cite{rosenthal}  &$13a$ & $13a$ & $-$ &  \\
\hline
FeSe~\cite{song, Song:2011aa,uday} & $-$ & $-$ &  $16a$ & \\ 
[1ex] 
\hline\hline 
\end{tabular}
\vspace{-2mm}
\end{table} 
%
\\

Finally, we should note that if the OS between  the $d_{xz}$ and $d_{yz}$ orbitals  is reversed ($\delta\rightarrow -\delta$), 
the QPI patterns in the pure SC state   or in the nematic state gets rotated by  $\pi/2$ for the same energy.
 However, its effect   is nontrivial in the SDW state because it involves a significant reconstruction of the electronic structure. 
 Also, we find that the patterns loose the 1d characteristics  which is otherwise  strongly favored  when the orbital $d_{xz}$ is lower in energy (see supplementary material for more details). 
\\

To conclude, the occurrence of parallel rod-like 
structures in the momentum space QPI 
or 1d spatial  modulation of the LDOS
in various phases of pnictides and 
chalcogenides is an indication of common
factor at play. 
We identify this as a symmetry breaking term
involving non degenerate
$d_{xz}$ and $d_{yz}$ orbitals. Incorporating such
a term while considering different 
phases, we have obtained all the essential features of QPI 
patterns and particularly the 1d LDOS modulations. 
In addition, we find it crucial that the
energy of $d_{xz}$ orbital be lower so that the 
orientation of anisotropic structures is robust against the change in 
quasiparticle energy. It is also 
illustrated how the non-uniform spectral-density
distribution along the 
constant energy contours, because of the nearby 
band extrema, leads to 
highly anisotropic impurity scattering. 
\\

{\it Acknowledgements:}
We are grateful to P. Wahl, S. Wirth, S. R\"o{\ss}ler, S. Borisenko, Z. Sun, Y. Bang and
I. Eremin for fruitful discussions.
We acknowledge the use of
 HPC cluster at HRI. 
 A.A. acknowledges support
 through National Research Foundation of Korea (NRF) funded by the Ministry of Science of Korea (Grants No. 2015R1C1A1A01052411 and No. 2017R1D1A1B03033465), and by the National Foundation of Korea (NRF) funded by the Ministry of Science, ICT and Future Planning (No. 2016K1A4A4A01922028).

\bibliography{References}



\clearpage
\newpage

\appendix

\graphicspath{{Figs_Sup/}}


\section*{supplementary Material}

\subsection{Quasiparticle interference in the superconducting state}
Modulation in the DOS due to an impurity atom is given 
by 
\be
\delta \rho({\bf q},\omega) = \frac{i}{2\pi} \sum_{\k}
g(\k, {\bf q},\omega),
\ee
where $g(\k, {\bf q},\omega)$ in terms of the change in the Green's function is
\be
g(\k, {\bf q},\omega) =  \sum_{{\rm i} \le5} (\delta {G}^{\rm ii}(\k,{\bf k^{\prime}},
\omega)-  \delta {G}^{\rm ii*}({\bf k^{\prime}},\k,\omega)).
\ee
Here, $\q= \k-\k^{\prime} $.
The change in the Green's function matrix due
to a single non-magnetic 
impurity is 
\be
\delta \hat{G}(\k, \k^{\prime}, \omega) = \hat{G}^0(\k,\omega) 
\hat{T}(\omega) \hat{G}^0(\k^{\prime},\omega),
\ee
where the meanfield Green's function is given by  
$$
\hat{G}^0(\k,\omega) = 
[
(\omega+i\eta)\hat{\bf I}- \hat{\cal H}(\k)
]^{-1},
$$
and the t-matrix is obtained as
\be
\hat{T}(\omega) = (\hat{\bf 1} - \hat{V} \hat{\mathcal{G}}(\omega))^{-1}\hat{V},
\ee
and $\hat{\bf I}$ represents  a 10$\times$10 identity 
matrix. 
Furthermore, we define 
\be
\hat{\mathcal{G}}(\omega) = \frac{1}{N} \sum_{\k} \hat{G}^{0}(\k, \omega).
\ee
and
\be
\hat{V} = V_{\rm imp}
\begin{pmatrix} 
\hat{\bf 1} & \hat{{\rm O}} \\
\hat{{\rm O}} & - \hat{\bf 1}  
\end{pmatrix},
\ee
where  $\hat{\bf 1}$ and $\hat{{\rm O}}$ are the  5$\times$5 identity 
matrix and null matrices, respectively.
Finally,  LDOS modulation or 
QPI in real space,
$\delta \rho({\bf r},\omega)$,
can be obtained by Fourier transform 
of $\delta \rho({\bf q},\omega)$.

\begin{figure}[t]
\begin{center}
\hspace{0.1mm}
\psfig{figure=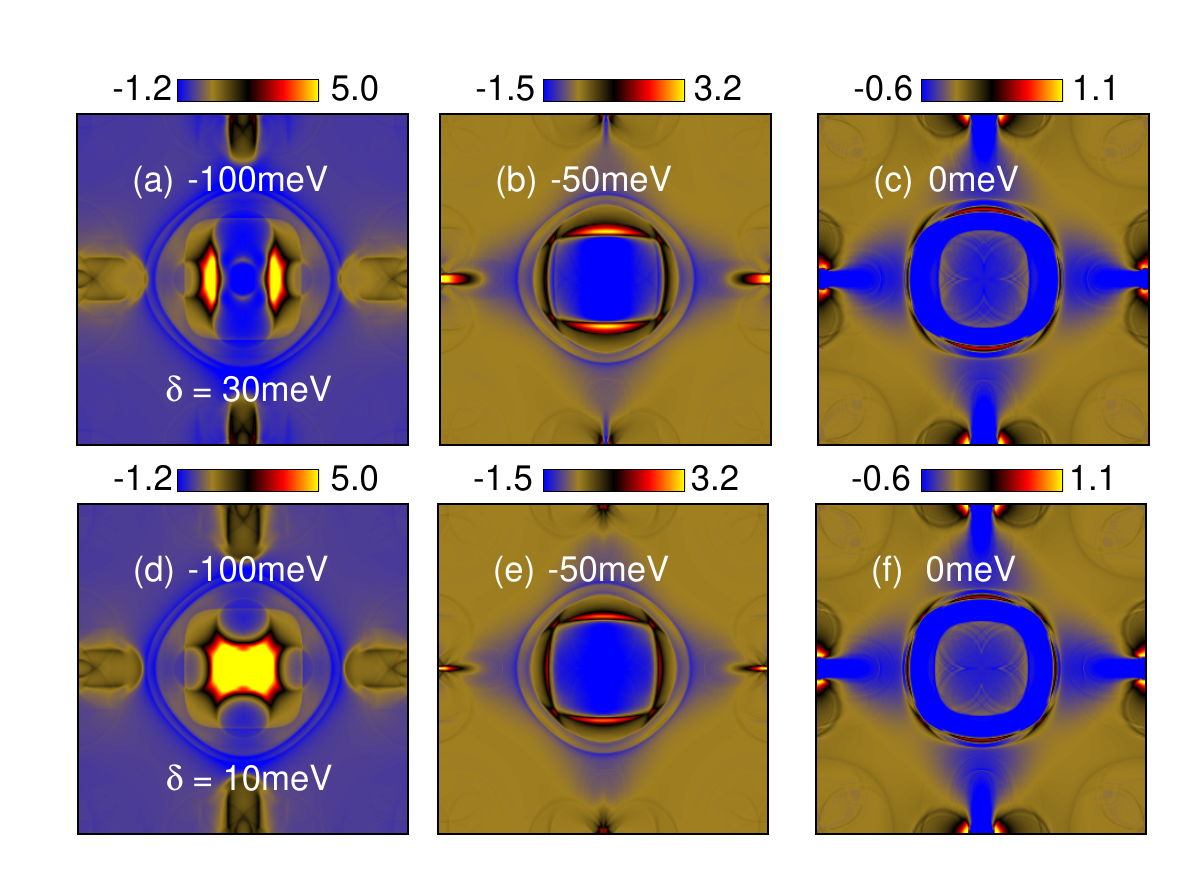,width=\linewidth}
\end{center}
\vspace{-6mm}
\caption{Momentum-space quasiparticle interference patterns in the nematic phase with orbital-splitting $\delta = 30$meV (first row (a), (b) and (c)) and 10meV (second row (d), (e) and (f)). Although features corresponding to one-dimensional LDOS modulation weaken on decreasing $\delta$ but still noticeable even for moderate value of $\delta = 30$meV.}
\label{sup1}
\end{figure}
\begin{figure}[t]
\begin{center}
\hspace{-2.5mm}
\psfig{figure=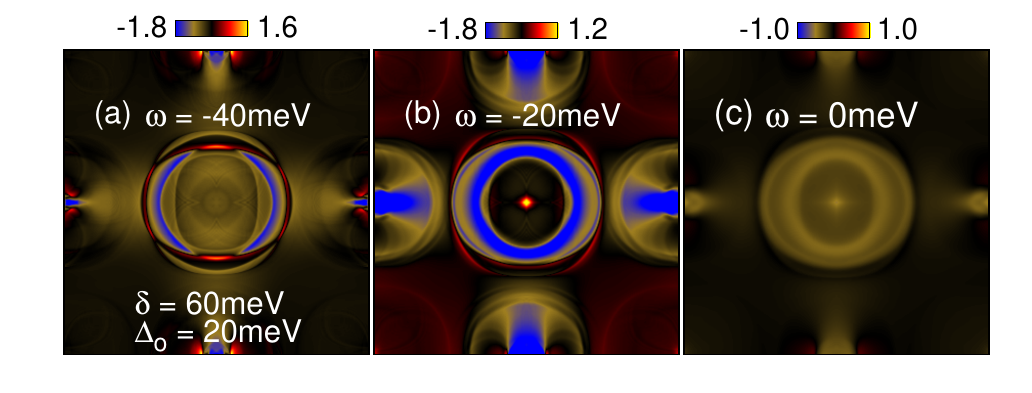,width=\linewidth}
\end{center}
\vspace{-6mm}
\caption{Quasiparticle interference patterns in the superconducting  state for $\omega = -40$meV, (b) -20meV (c) 0meV.}
\label{sup2}
\end{figure}


\subsection{Quasiparticle interference in the spin-density wave state}
As we descussed in the main text, the MF Hamiltonian in the  SDW state is obtained as 
\be
  {\mathcal{H}}_{SDW}  = 
\sum_{{\bf k }\sigma}\Psi'^{\dagger}_{{\bf k} \sigma}
\begin{pmatrix}
{\hat{\varepsilon}_{\k}+\hat{N}}  & {{\rm sgn}\bar{\sigma}\hat{W}} \\
{{\rm sgn}\bar{\sigma}\hat{W}} & {\hat{\varepsilon}_{\bf {k+Q}}+\hat{N}}
\end{pmatrix}
\Psi'_{{\bf k} \sigma}.
\ee
The matrix elements of matrices,
$\hat{W}$ and $\hat{N}$, in the above equation are defined as 
\bea
2W_{\mu\mu} &=& Um_{\mu\mu}+J\sum_{\mu \ne \nu}m_{\nu\nu} \nonumber\\
2W_{\mu\nu} &=& Jm_{\mu\nu}+(U-2J)m_{\nu\mu}
\eea
and 
\bea
2N_{\mu\mu} &=& Un_{\mu\mu}+(2U-5J)\sum_{\mu \ne \nu}n_{\nu\nu} \nonumber\\
2N_{\mu\nu} &=& Jn_{\mu\nu}+(4J-U)n_{\nu\mu},
\eea
where charge densities and magnetizations are given by
\be
n_{\mu\nu} = \sum_{\k \sigma} \langle d^{\dagger}_
{\k \mu \sigma}d_{\k \nu \sigma}\rangle,  \,\,\, 
m_{\mu\nu} = \sum_{\k \sigma} \langle d^{\dagger}_
{\k \bar{\mu} \sigma}d_{\k \nu \sigma}\rangle .
\ee
Note that ${d}^{\dagger}_{{\k}\bar{\mu}\sigma}$ = 
$d^{\dagger}_{{\k+\Q}\mu\sigma}$ with bar over 
orbital indices indicates shifting of momentum by ${\bf Q} 
= (\pi, 0)$. Summation over $\k$ is in 
the first Brillouin zone.

\begin{figure}[t]
\begin{center}
\hspace{-1.5mm}
\psfig{figure=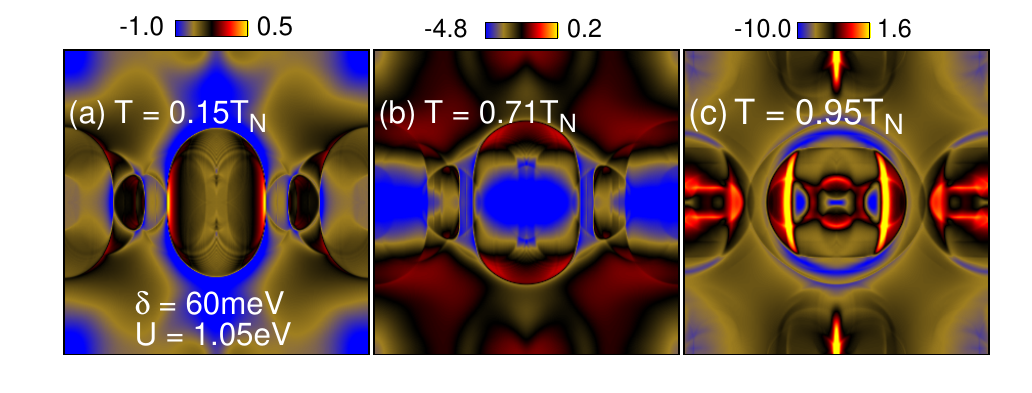,width=\linewidth}
\end{center}
\vspace{-6mm}
\caption{Quasiparticle interference patterns in the ($\pi$, 0) SDW state obtained for $U = 1.05$eV, $J$ = 0.25$U$ whereas orbital-splitting $\delta = 60$ meV. Total magnetization at $T \approx 0$K is $m \approx 0.8$.  }
\label{sup3}
\end{figure}


\begin{figure}[t]
\begin{center}
\hspace{-1.5mm}
\psfig{figure=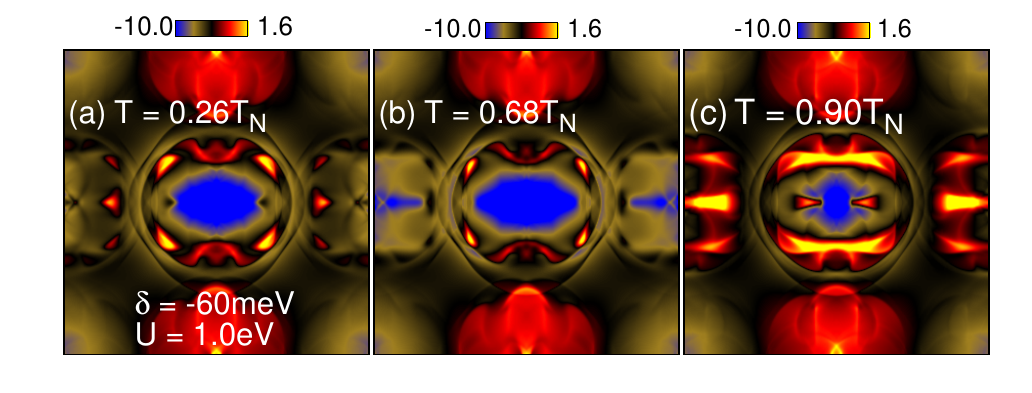,width=\linewidth}
\end{center}
\vspace{-6mm}
\caption{Quasiparticle interference patterns in the ($\pi$, 0) SDW state with sign of splitting reversed $\delta = -60$meV. Interaction parameter $U = 1.0$eV.}
\label{sup4}
\end{figure}
The expressions remains similar to the
case of superconducting  state. However, there are 
several differences as well. Impurity matrix is now 
\be
\hat{V} = V_{\rm imp}
\begin{pmatrix} 
\hat{\bf 1} & \hat{\bf 1} \\
\hat{\bf 1} & \hat{\bf 1}
\end{pmatrix}.
\ee
The change in the DOS is given 
by $\delta \rho_{\alpha} ({\bf q},\omega)$
\be
\delta \rho_{\alpha}({\bf q},\omega) = \frac{i}{2\pi} 
\sum_{\k} g_{\alpha}(\k, {\bf q},\omega)
\ee
with
\bea
g_0(\k, {\bf q},\omega) &=&  \text{Tr} \delta 
\hat{G}(\k,{\bf k^{\prime}},\omega)- 
\text{Tr} \delta \hat{G}^*({\bf k^{\prime}},\k,\omega) \nonumber\\
g_1(\k, {\bf q},\omega) &=&   \sum_{\mu\le5}\delta {G}_{\mu,\mu+5}
(\k,{\bf k^{\prime}},\omega)- 
\delta {G}^*_{\mu,\mu+5}({\bf k^{\prime}},\k,\omega) \nonumber\\
g_2(\k, {\bf q},\omega) &=&   \sum_{\mu\le5}\delta {G}_{\mu+5,\mu}
(\k,{\bf k^{\prime}},\omega)- 
\delta {G}^*_{\mu+5,\mu}({\bf k^{\prime}},\k,\omega)  \nonumber\\
\eea
and corresponding LDOS modulation is obtained as
\bea
\delta \rho({\bf r}_{\rm i},\omega) &=& \frac{1}{ N} \sum_{\q}
\Big[
 \delta\rho_0(\q, \omega)e^{i\q\cdot{\bf r}_{\rm i}} +
 \delta\rho_{1}(\q, \omega) e^{i(\q-\Q)\cdot{\bf r}_{\rm i}} \nonumber\\
 &&\hspace{1.2cm}+ \delta\rho_{2}(\q, \omega) e^{i(\q+\Q)\cdot{\bf r}_{\rm i}}
 \Big].
\eea

We have also examined the QPI patterns in the nematic phase for smaller value of orbital splitting $\delta$. Although momentum space QPI features corresponding to quasi-one dimensional LDOS modulation persist even for smaller $\delta$, it does weaken continuously on decreasing the latter (Fig. \ref{sup1}). This is not surprising because the asymmetry in the quasiparticle spectrum associated with the breaking of four-fold rotational symmetry decreases with $\delta$. 

Fig. \ref{sup2} shows QPI patterns in the SC state. It can be clearly seen that when $\omega$ decreases and approaches $\Delta_0 =20$meV, one dimensional characteristics declines continuously until it is lost completely when $\omega \lesssim \Delta_0$.

The QPI patterns are very sensitive to the size of the magnetic moment in the SDW state. To illustrate this, we have calculated the patterns for $U = 1.05$eV when $\delta = 60$meV. The net magnetic moment in the self-consistently obtained SDW state is $m \approx 0.8$ for temperature close to 0K. The results are shown in Fig. \ref{sup3}. Features corresponding to quasi one dimensional modulation in the LDOS can be seen just below the Neel temperature $ T_N$ where the band reconstruction is insignificant and the magnetic moment is small $m \approx 0.1$. However, the same is not true for further lower temperatures such as $T \sim 0.7T_N$ or $0.15T_N$, where magnetic moments are larger $m \sim 0.6$ or $0.8$. Thus, when the magnetic exchange splitting $2\Delta \gtrsim \delta$, a complex band reconstruction perhaps yields a modification in the anisotropy in such a way that one-dimensional features are lost. Note that the magnetic moments are usually smaller than $1$ in most of the pnictides \cite{cruz}.

Throughout the calculations, the orbital splitting was such that $d_{xz}$ was lower in energy. We have investigated the effect of orbital splitting with sign reversed, {\it i.e}, when $d_{yz}$ orbital is lower in energy instead. The patterns in the superconducting state or in the nematic state gets rotated by 90$^{\circ}$ for the same energy if the sign of $\delta$ is reversed. This is because $d_{xz}$ is mapped to $d_{yz}$ by a 90$^{\circ}$ rotation. The effect, however, is non-trivial in the SDW state for the reason that the four-fold rotation symmetry is also broken by the SDW state so that $d_{xz}$ and $d_{yz}$ orbitals are unequally populated. Therefore, the role of additional term which removes the degeneracy of $d_{xz}$ and $d_{yz}$ is not itself clear. We examined the QPI patterns in the ($\pi$, 0) SDW state with $\delta =$ -60meV, where $U = 1.0$eV.   Clearly, the patterns loose one-dimensional characteristics. In other words, one-dimensional QPI patterns in the SDW state with ordering wave-vector ($\pi, 0$) is supported only when the orbital $d_{xz}$ is lower in energy.

\end{document}